\documentclass[a4paper,12pt]{article}
\usepackage{graphicx, color}
\usepackage{latexsym}
\let\section=\subsection     \let\subsection=\subsubsection                
\begin{document}
\def\BA{\Bbb A}
\def\BC{\Bbb C}
\def\BR{\Bbb R}
\newfont{\titlefont}{cmssbx10 scaled\magstep5}
\newcommand{\CA}{{\cal A}}
\newcommand{\CB}{{\cal B}}
\newcommand{\CC}{{\cal C}}
\newcommand{\CD}{{\cal D}}
\newcommand{\CE}{{\cal E}}
\newcommand{\CF}{{\cal F}}
\newcommand{\CG}{{\cal G}}
\newcommand{\CH}{{\cal H}}
\newcommand{\CI}{{\cal I}}
\newcommand{\CJ}{{\cal J}}
\newcommand{\CK}{{\cal K}}
\newcommand{\CL}{{\cal L}}
\newcommand{\CM}{{\cal M}}
\newcommand{\CN}{{\cal N}}
\newcommand{\CO}{{\cal O}}
\newcommand{\CP}{{\cal P}}
\newcommand{\CQ}{{\cal Q}}
\newcommand{\CR}{{\cal R}}
\newcommand{\CS}{{\cal S}}
\newcommand{\CT}{{\cal T}}
\newcommand{\CU}{{\cal U}}
\newcommand{\CV}{{\cal V}}
\newcommand{\CW}{{\cal W}}
\newcommand{\CX}{{\cal X}}
\newcommand{\CY}{{\cal Y}}
\newcommand{\CZ}{{\cal Z}}
\newcommand{\bea}{\begin{eqnarray}} \newcommand{\eea}{\end{eqnarray}}
\newcommand{\beqa}{\begin{eqnarray}} \newcommand{\eeqa}{\end{eqnarray}}
\newcommand{\beq}{\begin{equation}} \newcommand{\eeq}{\end{equation}}
\newcommand{\non}{\nonumber} \newcommand{\eqn}[1]{\beq {#1}\eeq}
\newcommand{\eu}{Euclidean\ }
\newcommand{\lmk}{\left(} \newcommand{\rmk}{\right)}
\newcommand{\lkk}{\left[}
\newcommand{\rkk}{\right]} \newcommand{\lhk}{\left \{ }
\newcommand{\rhk}{\right \} } \newcommand{\lnk}{\left \{ }
\newcommand{\rnk}{\right \} } \newcommand{\del}{\partial}
\newcommand{\abs}[1]{\vert{#1}\vert}
\newcommand{\vect}[1]{\mbox{\boldmath${#1}$}} \newcommand{\bib}{\bibitem}
\newcommand{\new}{\newblock}
\newcommand{\la}{\left\langle} \newcommand{\ra}{\right\rangle}
\newcommand{\bfx}{{\bf x}} \newcommand{\bfk}{{\bf k}}
\newcommand{\vex}{{\vect x}}
\newcommand{\vek}{{\vect k}} \newcommand{\vep}{{\vect p}}
\newcommand{\veq}{{\vect q}} \newcommand{\vev}{{\vect v}}
\newcommand{\vej}{{\vect j}} \newcommand{\veg}{{\vect \gamma}}
\newcommand{\vena}{{\vect \nabra}} \newcommand{\vebt}{{\vect \beta}}
\newcommand{\gtilde}{~\mbox{\raisebox{-1.0ex}{$\stackrel{\textstyle >}
{\textstyle \sim}$ }}}
\newcommand{\ltilde}{~\mbox{\raisebox{-1.0ex}{$\stackrel{\textstyle <}
{\textstyle \sim}$ }}}
\newcommand{\gsim}{~\mbox{\raisebox{-1.0ex}{$\stackrel{\textstyle >}
{\textstyle \sim}$ }}}
\newcommand{\lsim}{~\mbox{\raisebox{-1.0ex}{$\stackrel{\textstyle <}
{\textstyle \sim}$ }}}
\newcommand{\mh}{m_H} \newcommand{\veff}{V_{\rm eff}}
\newcommand{\phip}{\phi_+}
\newcommand{\eff}{\rm eff}
\newcommand{\Ghat}{\hat{\Gamma}}
\newcommand{\mpl}{M_{Pl}}
\newcommand{\gn}{g_{*N}}
\begin{center}
%
   {\large \bf  BARYON DIFFUSION CONSTANT }\\[2mm]
   {\large \bf IN HOT AND DENSE HADRONIC MATTER }\\[2mm]
   {\large \bf BASED ON AN EVENT GENERATOR URASiMA}\\[5mm]
N.~SASAKI, O.~MIYAMURA, S.~MUROYA$^{*}$ and C.~NONAKA \\[5mm]
{\small \it  Department of Physics, Hiroshima University, \\
{\it Higashi-Hiroshima 739-8526, Japan }\\
{\it $^*$Tokuyama Women's College, Tokuyama, 745-8511, Japan } \\[8mm] }
\end{center}

\begin{abstract}
We generate the statistical ensembles in equilibrium 
with fixed temperature and chemical
potential by imposing periodic boundary condition to the
simulation of URASiMA(Ultra-Relativistic AA collision Simulator based 
on Multiple Scattering Algorithm).
By using the generated ensembles, we investigate the temperature dependence
and the chemical potential dependence of the nucleon diffusion constant of a
dense and hot hadronic matter.
\end{abstract}

\section{ Introduction} \label{intr}

Because of the highly non-perturbative property of a hot and dense hadronic
state, the thermodynamical properties and transport
coefficients has been hardly investigated. 
In this paper\cite{muroya}, we evaluate the transport coefficients by using statistical
ensembles generated by Ultra-Relativistic A-A collision simulator based on
Multiple Scattering Algorithm (URASiMA). Originally, URASiMA is an event
generator for the nuclear collision experiments based on the Multi-Chain
Model(MCM) of the hadrons\cite{Kumagai}.
Some of us(N.\ S.\ and O.\ M.) has already discussed thermodynamical properties
of a hot-dense hadronic state based on a molecular dynamical simulations of
URASiMA with periodic condition\cite{Sasaki1,RQMD}.  
We improve URASiMA
to recover detailed balance at temperature below two hundred MeV.
As a result, Hagedorn-type behavior in
the temperature disappears\cite{Sasaki2}. This is the first calculation of
the transport coefficient of a hot and dense hadronic matter based on
an event generator.

\section{URASiMA for Statistical Ensembles} \label{URA}

In order to obtain equilibrium state, we put the system in a box and impose
a periodic condition to URASiMA as the space-like boundary condition. Initial
distributions of particles are given by uniform random distribution of
baryons in a phase space. Total energy and
baryon number in the box are fixed at initial time and conserved through-out
simulation. Running URASiMA many times with the same total energy
and total baryons in the box and taking the stationary configuration later
than $t=150$ fm/c, we obtain statistical ensemble with fixed
temperature and fixed baryon number(chemical potential).  By using the 
ensembles obtained through above mentioned manner, we can
evaluate thermodynamical quantities and equation of states\cite{Sasaki2}.

\section{Transport Coefficients} \label{Trans}

According to the Kubo's Linear Response Theory, the correlation of the
currents stands for the admittance of the system(first fluctuation dissipation
theorem) and equivalently, random-force correlation gives the impedance(Second
fluctuation dissipation theorem) \cite{Kubo}. As the simplest example, we
here focus our discussion to the diffusion constant. First
fluctuation dissipation theorem tells us that diffusion constant $D$ is
given by current(velocity) correlation,
\beq
D =\frac{1}{3} \int_{0}^{\infty}<\vev(t)\cdot \vev(t+t')> dt'. \label{eqn;fdt} \eeq
Average $<\cdots>$ is given by,

\beq
<\cdots> = \frac{1}{\mbox{number of ensembles}}\sum_{\mbox{ensemble}}
\label{eqn;avg}
\frac{1}{\mbox{number of particle}}\sum_{\mbox{particle}} \cdots . \eeq
Figure 1 shows correlation function of the velocity of baryons. The figure
indicates that exponential damping is very good approximation.
In the case that the correlation decrease exponentially, i.e., \beq
<\vev(t)\cdot \vev(t+t')> \propto exp{(- \frac{t'}{\tau})}, \label{eqn;rlx}
\eeq
with $\tau$ being relaxation time,
diffusion constant can be rewritten in the simple form, \beq
D = \frac{1}{3} <\vev(t)\cdot \vev(t)> \tau .\label{eqn;difcon} \eeq
\begin{center}
\begin{minipage}{13cm}
\baselineskip=12pt
  \includegraphics[scale=0.7]{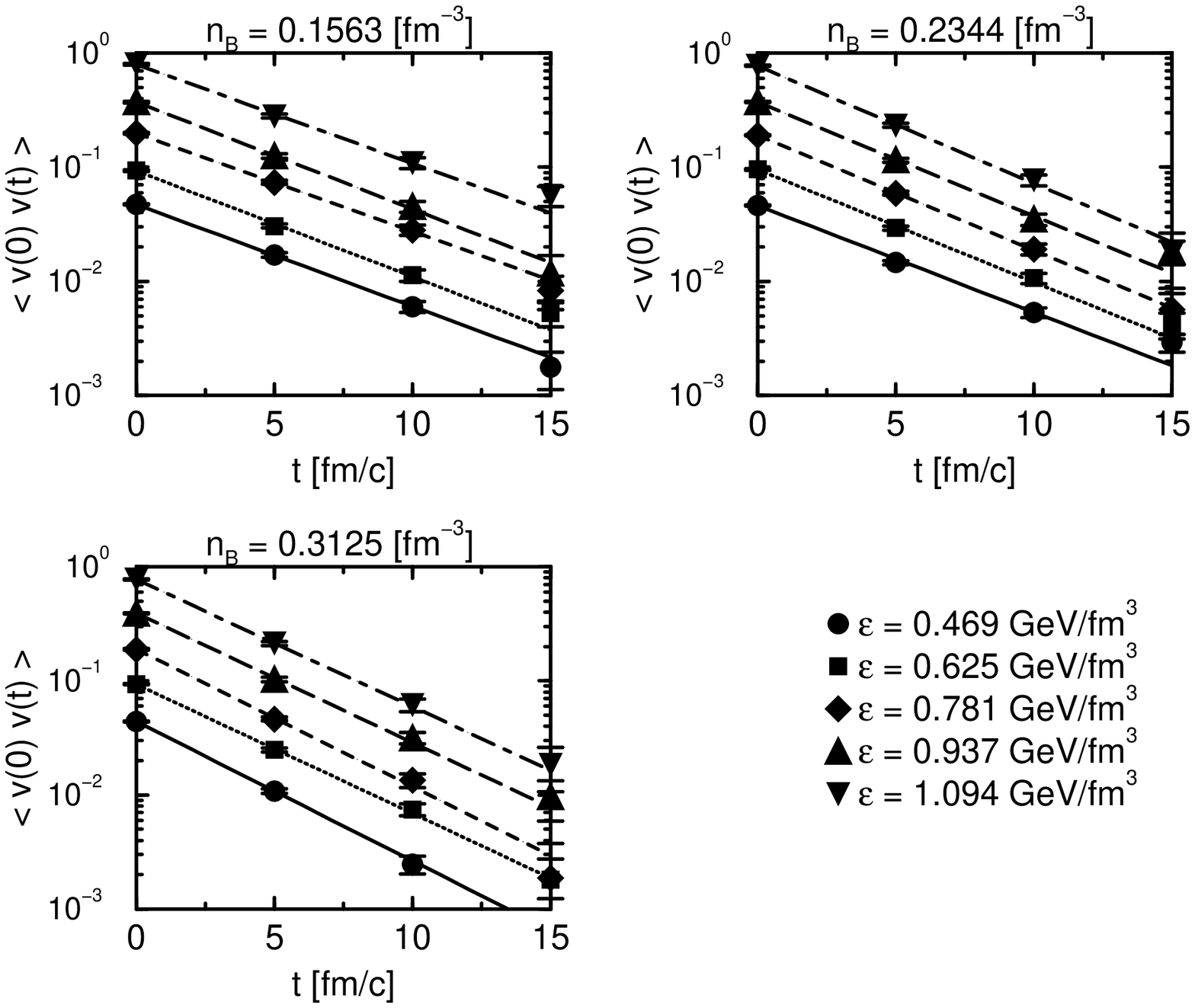}\\
{ \begin{small}
Fig.~1. Velocity correlation of the baryons as a function of time.
    Lines correspond to the fitted results by exponential function.
    Normalizations of the data are arbitrary.
 \end{small}
}
\end{minipage}
\end{center}

Usually, diffusion equation is given as, \beq
{\frac{\partial}{\partial t}} f(t,\vex) = D \nabla^2 f(t,\vex),
\label{eqn;difeq}
\eeq
and diffusion constant $D$ has dimension of $[L^2 /T].$ Because of
relativistic nature of our system, we should use $\vebt = \frac{\vev}{c} =
\frac{\vep}{E}$ instead of $\vev$ in eq.(\ref{eqn;fdt}) and $D$ is obtained
through,

\beqa
D &=& \frac{1}{3}\int_{0}^{\infty}<\vebt(t)\cdot \vebt(t+t')> dt' c^2 .\label{eqn;fdt2}
\\
&=&\frac{1}{3}<\vebt(t)\cdot \vebt(t)> c^2 \tau .\label{eqn;difcon2} \\
&=& \frac{1}{3}<\left(\frac{\vep(t)}{E(t)}\right)\cdot
\left(\frac{\vep(t)}{E_(t)}\right)> c^2 \tau \eeqa
with $c$ being the velocity of light.
Figure 2 displays the our results of baryon diffusion constant in a hot and
dense hadronic matter.
\begin{center}
\begin{minipage}{13cm}
\baselineskip=12pt
  \includegraphics[scale=0.50]{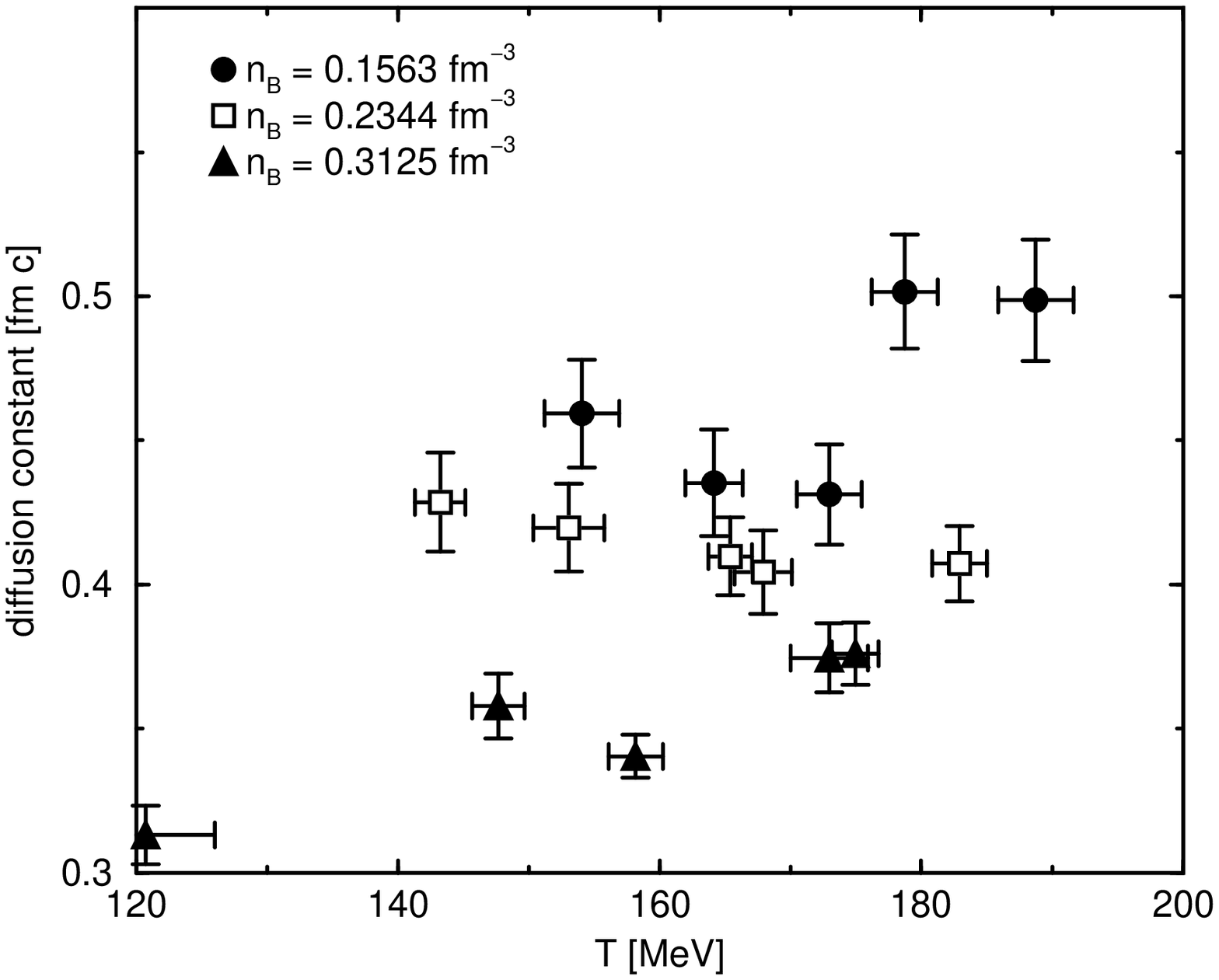}\\
{\begin{small}
Fig.~2. Diffusion constant of baryons.
\end{small}
}
\end{minipage}
\end{center}
Our results shows clearer dependence on the baryon number density while
dependence on energy density is mild. This results means importance of
baryon-baryon collision process for the random walk of the baryons
and thus non-linear diffusion process of baryons occurs. In this sense, we
can state that
baryon number density in our system is still high. In the inhomogeneous
big-bang nucleosynthesis scenario, baryon-diffusion play an important roll.
The leading part of the scenario is played by the difference
between proton diffusion and neutron diffusion\cite{inho}. In our
simulation,
strong interaction
dominates the system and we assume charge independence in the strong
interaction, hence, we can not discuss difference between proton and
neutron.
However obtained diffusion constant of baryon in our simulation can give
some
kind of restriction to the diffusion constants of both proton and neutron.

Because fundamental system in URASiMA is high energy
hadronic collisions, we use relativistic notations usually. However,
diffusion equation
(\ref{eqn;difeq}) is not Lorentz's covariant and is available only on the
special system i.e. local rest frame of the thermal medium. For the
full-relativistic description of the space-time evolution of a hot and
dense matter, we need to establish relativistic Navier-Stokes
equation\cite{Namiki}.
And taking correlation of appropriate currents, we can easily evaluate
viscosities and heat conductivity in the same manner \cite{Sasaki3} .

\section{Concluding Remarks} \label{conc}

Making use of statistical ensembles obtained by an event generator URASiMA,
we evaluate diffusion constants of baryons in the hot and dense hadronic
matter. Our results show strong dependence on baryon number density and
weak dependence on temperature.
The temperature in our simulation is limited only small range, i.e., from
100 MeV to 200 MeV, and this fact can be one of the reasons  why the change
of
diffusion constant of temperature is not clear.
Strong baryon number density dependence indicates that, for the baryon
diffusion process, baryon plays more important roll than light mesons. In this
sense our simulation corresponds to high density region and non-linear
diffusion process occurs. Calculation of the diffusion constants is the
simplest examples of first fluctuation dissipation theorem.
In principle, taking correlation of appropriate currents, i.e. energy
flow, baryon number current, stress-tensor, etc.,
we can evaluate any kinds of transport
coefficients.


\end{document}